\pdfoutput=1
\documentclass[bibnotes,prl,amsmath,citeautoscript,twocolumn,
               preprintnumbers,amssymb,amsmath,floatfix,
               letterpaper]{revtex4}
\usepackage{graphicx} 
\usepackage{microtype}

\begin{document}
\title{Substrate coherency-driven octahedral rotations in perovskite oxide films}
  \author{James M.\ Rondinelli}
     \email[Address correspondence to: ]{rondo@mrl.ucsb.edu}
  \altaffiliation{Current address: X-ray Science Division, Argonne National Laboratory, Argonne, Illinois 60439, USA}
  \author{Nicola A.\ Spaldin}
  \affiliation{Materials Department, University of California, Santa Barbara, 
	       CA, 93106-5050, USA}
\date{\today}

\begin{abstract}
\sloppy
We perform first-principles density functional calculations to 
explore the role of substrate proximity effects on the octahedral rotation 
patterns in perovskite oxide superlattices.
With cubic perovskite SrFeO$_3$ as our model film and tetragonal SrTiO$_3$ as 
the substrate, we show that in most cases the substrate octahedral rotation 
patterns propagate into the film across the heterointerface.
We also identify elastic boundary conditions for which the enforced structural 
coherence induces atomic displacement patterns that 
are not found in the bulk phase diagram of either individual constituent.
We suggest that such substrate coherency-induced octahedral texturing of thin 
film oxides is a promising approach for tuning the electronic structure of functional 
oxide thin films.
\end{abstract}

\maketitle

\sloppy
The use of substrate-induced bi-axial strain to modify the properties 
of epitaxial thin films has been demonstrated for a wide range of 
phenomena and materials, including mobility in semiconductors, 
Curie temperatures in ferromagnets and ferroelectric polarizations in 
complex oxides \cite{Schlom/Triscone_et_al:2007}.
While the main effect of strain in strongly covalently bonded materials
such as semiconductors is to change the bond lengths, the flexible 
corner-sharing networks of oxygen polyhedra in complex oxides 
provide additional routes for accommodating substrate-induced 
changes in lattice parameters. 
Established mechanisms include modifications of the oxygen polyhedral 
tilt patterns, and the formation of twin (change in orientation of long 
and short axes) or antiphase (variation in phase of polyhedral rotations) 
domains.
Recently it is been suggested that, in addition to changing the 
lattice parameter of a complex oxide film, the presence of a 
heterointerface could alter the relative stability of polyhedral tilting 
patterns in both the film and substrate through proximity effects  \cite{He/Xi_et_al:2004,Xie/Woicik_et_al:2008,Hoppler/Willmott:2008,Loetzsch/Seidel:2010}. 
While theoretical studies have shown that strain-induced competition 
between polyhedral rotation modes and other lattice distortions is 
crucial in determining the functional properties of complex oxides  
\cite{Okamoto/Millis/Spaldin:2006,Bhattacharjee/Bousquet/Ghosez:2009,Rondinelli/Eidelson/Spaldin:2009,Bousquet/Ghosez_et_al:2008,Hatt/Ramesh:2009},
very little is known about the extent to which substrate proximity 
modifies polyhedral tilt patterns. 
This is in part due to difficulties in obtaining high 
precision measurements of oxygen positions in superlattice and thin 
film interfaces \cite{Jia/Urban_et_al:2009,May/Rondinelli:2010}.
In this Letter, we use density functional theory (DFT) to calculate 
explicitly how the structural distortions of a substrate affect the atomic 
structure and properties of a coherent film. 
We take perovskite-structured SrFeO$_3$/SrTiO$_3$ as our model system, 
chosen for its continuous $A$-site sublattice, absence of polar discontinuity, 
and simple oxygen octahedral tilt patterns: 
SrFeO$_3$ has the ideal cubic $Pm\bar{3}m$ perovskite structure down to 
the lowest temperature studied (4~K) \cite{Macchesney/Potter_et_al:1965}, and 
the ground state $I4/mcm$ phase of SrTiO$_3$ (which is a widely 
used substrate) has a single octahedral instability with respect to the cubic phase \cite{Jauch/Palmer:1999} that condenses below $\sim$105~K. 
First, we investigate the effect of heterostructure periodicity 
in symmetric (SrTiO$_3$)$_n$/(SrFeO$_3$)$_n$, $n=1\ldots5$, 
and asymmetric  (SrTiO$_3$)$_n$/(SrFeO$_3$)$_m$, $n=1\ldots3,m=1\ldots3$ 
superlattices. 
We find that the octahedral rotations from the SrTiO$_3$ substrate 
propagate into the first two interfacial SrFeO$_3$ layers; for
highly confined ferrate layers  additional electronically-driven 
lattice instabilities occur that are not observed in bulk SrFeO$_3$.
Then we show that these octahedral and electronic lattice instabilities
cannot be induced in SrFeO$_3$ using bi-axial strain alone, indicating
that substrate coherency and confinement play a critical role in determining 
the heterointerface atomic structure.
Our DFT calculations are performed within the local 
spin-density approximation + Hubbard $U$ (LSDA$+U$) method 
using the Vienna {\it Ab initio} Simulation Package ({\sc vasp}) 
\cite{Kresse/Furthmueller_PRB:1996,Kresse/Joubert:1999}, with the Dudarev 
method \cite{Dudarev_et_al:1998} for the Hubbard correction to the LSDA 
exchange-correlation functional.
We use an effective $U$ parameter of 6~eV on the Fe $d$ orbitals,  
and impose ferromagnetic spin order in the SrFeO$_3$; 
this approach is consistent with earlier first-principles calculations \cite{Shein/Ivanovskii_et_al:2005}. 
Complete details of our calculations are reported elsewhere \cite{Rondinelli/Spaldin:2010}.
\begin{figure*} 
\centering
\includegraphics[width=0.83\textwidth]{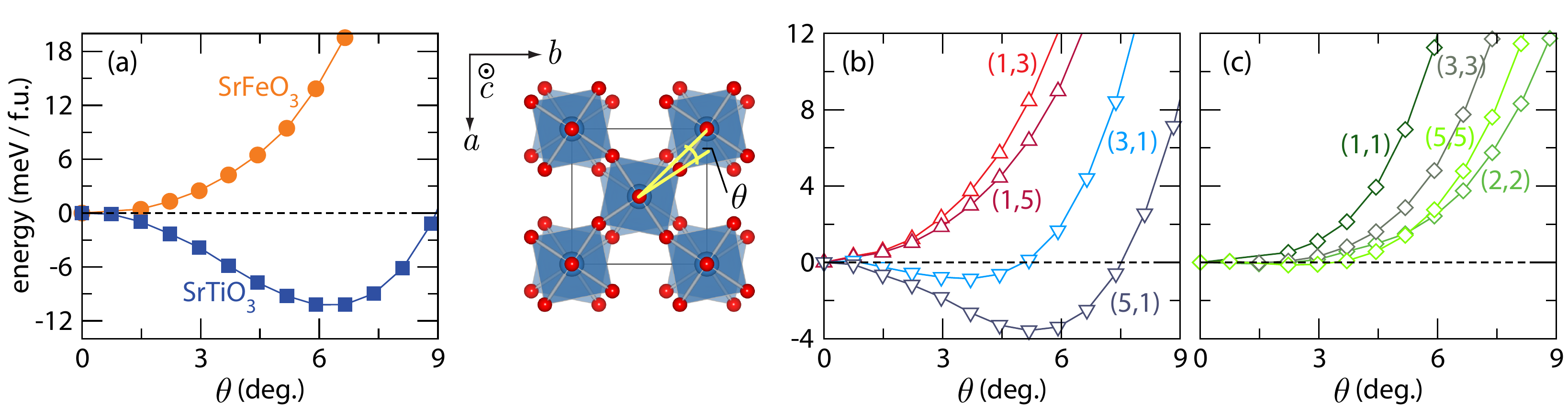}\vspace{-6pt}
\caption{\label{fig:homostrain}(Color) 
Energy versus rotation angle $\theta$ of the 
$a^0a^0c^-$ octahedral tilt for cubic SrFeO$_3$ and tetragonal SrTiO$_3$ (a).
The same mode homogeneously frozen into (b) asymmetric 
(SrTiO$_3$)$_n$/(SrFeO$_3$)$_m$ and (c) symmetric 
superlattices (right).}
\end{figure*}
We first calculate how freezing in the antiferrodistortive (AFD) 
$a^0a^0c^-$ octahedral tilt pattern of the low-temperature phase of 
SrTiO$_3$ \cite{Unoki/Sakudo:1967} affects the 
total energies of the two bulk materials which comprise the superlattices.
Figure \ref{fig:homostrain}(a) shows the calculated total energies 
as a function of increasing amplitude of octahedral rotation angle $\theta$,
referenced to the un-rotated states at the L(S)DA equilibrium lattice parameters.
As expected for bulk SrFeO$_3$, we find that the cubic perovskite 
structure is stable with respect to the octahedral rotation mode.
In contrast, the $a^0a^0c^-$ tilt pattern is energy lowering  
for tetragonal SrTiO$_3$, consistent with the low-temperature experimental 
structure\footnote{%
We note that there is an overestimate in our LDA calculation 
of the bulk rotation angle of $\theta=5.7^\circ$ compared to the 
experimental value of $\theta=2.1^\circ$ \cite{Courtens:1972}.
Previous work indicates that working at the experimental volume does 
not dramatically improve the calculated value since the 
inconsistency originates from quantum point fluctuations that 
are not captured by the exchange-correlation functionals available 
to DFT \cite{Sai/Vanderbilt:2000}.}.
We next explore whether heterostructuring 
SrFeO$_3$ layers with SrTiO$_3$ in different period superlattices 
induces AFD FeO$_6$ instabilities in the ferrate layers.
We examine symmetric and asymmetric superlattices, by stacking 
($n,m$) five-atom perovskite cells along the $c$-axis, with the
in-plane periodicity increased to be commensurate with the AFD
rotations.
We keep the in-plane lattice constant fixed to the LDA equilibrium 
value of 3.86~{\AA} to simulate epitaxial growth on tetragonal SrTiO$_3$ 
(SrFeO$_3$ is under a L(S)DA theoretical $+1.75$\% bi-axial strain). 
The out-of-plane lattice constant of each superlattice is then fully relaxed.
We first introduce the same octahedral rotation pattern as before uniformly 
into each layer of superlattice. 
The evolution in the total energy for the asymmetric ($n,m$) 
and symmetric ($n,n$) superlattices with increasing magnitude of  $\theta$ 
is shown in Figure \ref{fig:homostrain}(b) and (c).
For the asymmetric superlattices we find that when SrTiO$_3$ comprises the 
majority of the superlattice, a homogeneous rotation of all octahedra 
is favored over the unrotated configuration.
In contrast, the symmetric superlattices, and those with a greater fraction of 
SrFeO$_3$ layers are weakly, if at all, susceptible to the same rotational 
modes.
\begin{figure}[b]
\centering
\includegraphics[width=0.44\textwidth]{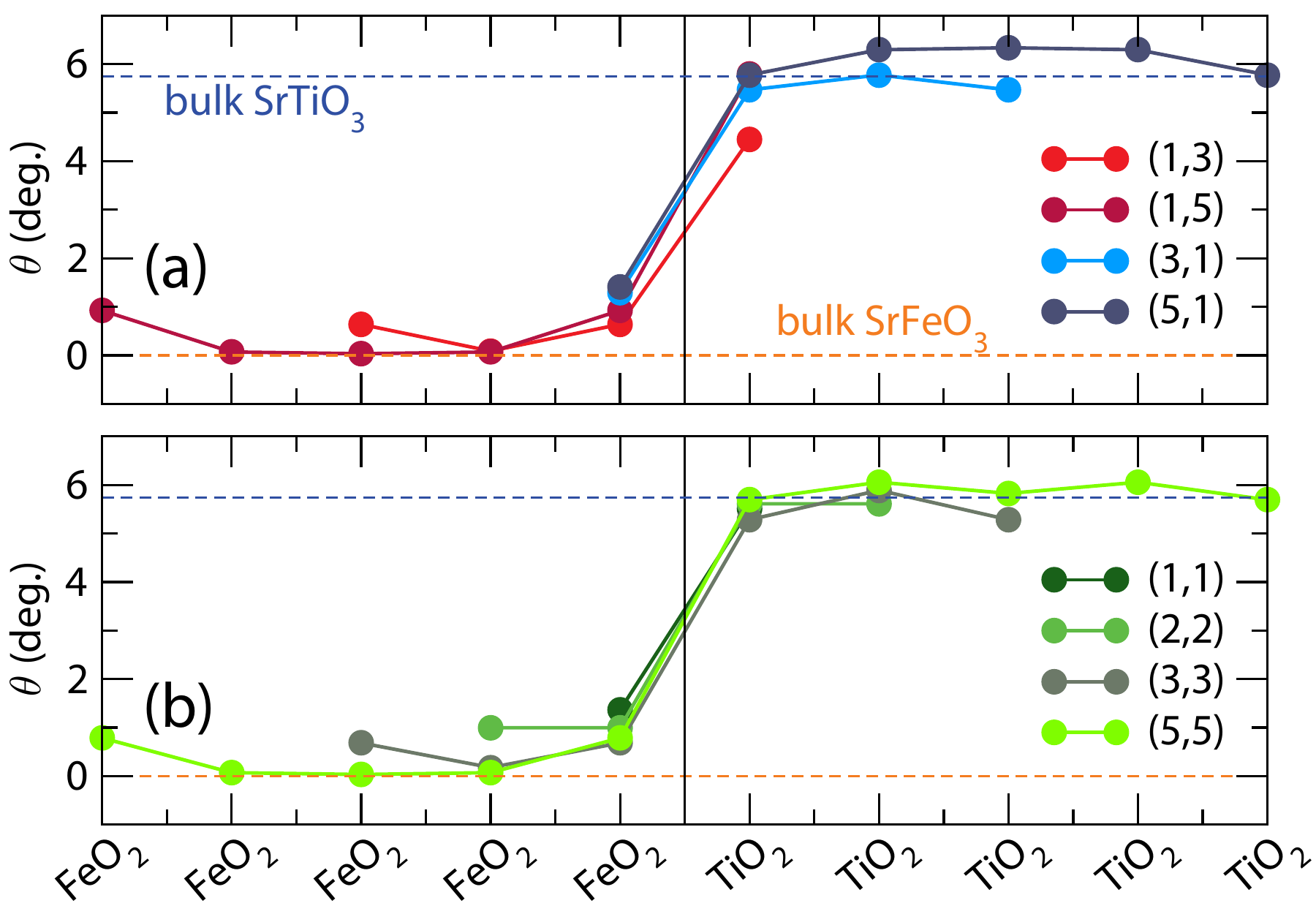}\vspace{-6pt}
\caption{(Color) 
The layer-by-layer resolved octahedral rotation angles ($\theta$) for 
the asymmetric (a) and symmetric (b) superlattices.
The magnitude of the SrTiO$_3$ AFD octahedral rotations rapidly 
decreases into the SrFeO$_3$ interfacial layers.
\label{fig:stfo} }%
\end{figure}
Since the homogeneous rotation throughout the superlattice is 
artificial, we next remove the homogeneity constraint 
and allow full relaxation of the oxygen positions in 
each layer within the symmetry of the $a^0a^0c^-$ tilt pattern.
We initialize the oxygen positions in each superlattice to configurations 
corresponding to the energy well minima shown in Figure \ref{fig:homostrain}.
(For homogeneous rotations that were found to be stable, we start with $\theta=2^\circ$.)
The results of the structural relaxation are shown in Figure \ref{fig:stfo}, 
where we plot the  layer-by-layer resolved rotations $\theta$ 
about the $c$ axis for the different period superlattices.
We find that in all cases, the octahedral rotations remain antiferrodistortive 
and that the SrTiO$_3$ substrate exhibits the  $a^0a^0c^-$ tilt pattern.
Interestingly, AFD octahedral rotations are induced in the 
interfacial SrFeO$_3$ layers, with  the magnitude of the octahedral 
rotations decaying exponentially from the interface into the center of 
the SrFeO$_3$ slab.
The layers in the center of each slab are close to their respective 
bulk calculated values, indicated by the broken lines, although
the highly asymmetric (5,1) superlattice shows a 9.3\%  
enhancement in $\theta$ at the center of the SrTiO$_3$ slab.
We next fully relax the atomic structure of each superlattice by removing the 
symmetry constraint imposed by the $a^0a^0c^-$ octahedral tilt pattern.
The asymmetric superlattices in which SrFeO$_3$ is the majority component do 
not show any considerable changes in the atomic displacements: there is only a 
small decrease in the magnitude of the $a^0a^0c^-$ octahedral tilt at 
the interface due to small changes in the apical cation-oxygen bond lengths.
We also found similar minor changes in the atomic structure for the symmetric 
superlattices ($n,n$) with $n\geq2$.
Therefore our earlier conclusion that only the first two 
ferrate interface unit cells are modified by the octahedral rotations 
found in the substrate remains for these cases.
Drastically different behavior is found, however, for superlattices 
with single unit cells of SrFeO$_3$: in the 
($1,1$), ($3,1$) and ($5,1$) superlattices  we find that, in addition 
to the original octahedral rotation pattern, the FeO$_6$ octahedra 
also exhibit Fe--O bond length distortions associated with 
electronic instabilities.
In each case, either a Jahn-Teller (JT) distortion, which produces two 
short and two long equatorial Fe--O bonds, or a ``breathing'' distortion 
that makes a uniform dilation and contraction of the Fe--O bonds is  
found to be stable.
Note that, for the single ferrate layer heterostructures, we obtain
JT- and breathing-distorted FeO$_6$ octahedra even when we disable the 
oxygen octahedral rotations in our calculations.
This indicates that these electronic instabilities are the consequence 
of the quantum confinement of the ferrate layer, which increases the
susceptibility to electron localization and in turn enhances orbital
degeneracy-lifting instabilities
\cite{Rondinelli/Spaldin:2010}.
In the unrotated case, the energies of the JT- and breathing-distorted 
structures are energetically equivalent within the resolution of our DFT 
calculations.
While the JT- and breathing distortions are not induced by the octahedral
rotations, when both rotations and JT- or breathing distortions are allowed,
we are able to resolve distinct ground states, in which the octahedral
rotations cooperate with a specific electronic instability.
In the (3,1) superlattice, for example, the ground state consists of
a large Jahn-Teller bond length distortion of 0.05~{\AA}, in 
combination with a completely different octahedral rotation pattern 
-- the well-known  perovskite $a^-a^-c^+$ GdFeO$_3$ tilt pattern --
that does not occur in either of the parent compounds.
%
%
If we instead enforce the usual $a^0a^0c^-$ tilt pattern of SrTiO$_3$, the 
superlattice exhibits a breathing distortion with Fe--O bond length 
differences of 6.6\% between the two inequivalent FeO$_6$ octahedral sites.
%

Finally, we check whether our finding of octahedral rotations within the
SrFeO$_3$ interfacial layers can be reproduced using strain alone, or
whether it requires substrate coherency. 
To explore the interaction between strain and the octahedral rotations on the 
SrFeO$_3$ layers we remove the SrTiO$_3$ substrate from our calculations 
and simulate homoepitaxially strained SrFeO$_3$, with the bi-axial constraint 
imposed by enforcing equal in-plane lattice parameters.
In addition to the ground state $a^0a^0c^-$ tilt pattern, we also examine 
the strain dependence of the in-phase $a^0a^0c^+$ rotation about the 
$c$-axis, the $a^-a^0c^0$  AFD tilt pattern about the $a$-axis that lies 
in the epitaxial plane, and the Jahn-Teller and breathing distortions.
%
%
Note that similar studies were performed previously for
SrTiO$_3$ \cite{Sai/Vanderbilt:2000,Lin/Guo:2006,Bousquet/Ghosez_et_al:2008}; 
we do not repeat those calculations here since our SrTiO$_3$ substrates 
are strain free.
\begin{figure}
\centering
\includegraphics[width=0.49\textwidth]{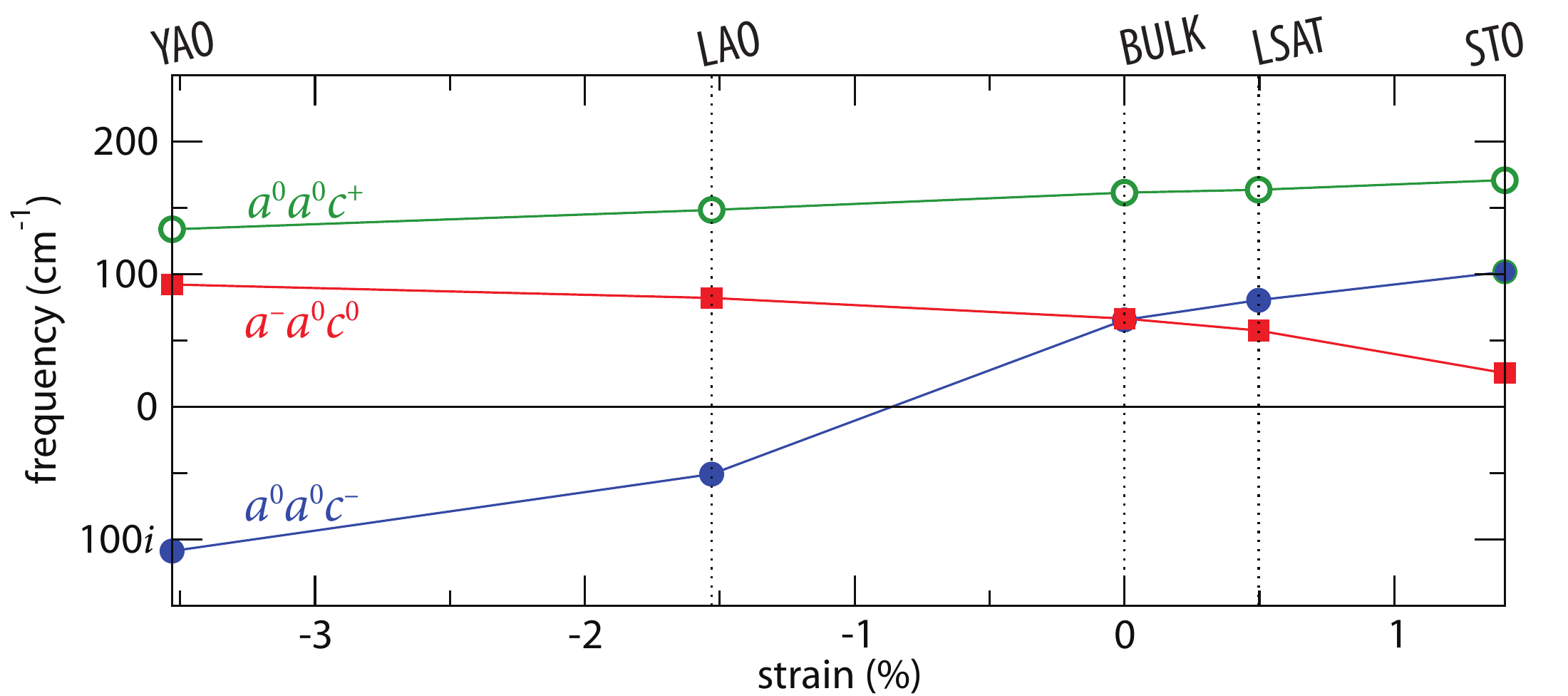}
\caption{\label{fig:sfo_strain} 
The effect of bi-axial strain on the mode stiffness of the rotational 
patterns in homogeneously strained SrFeO$_3$ films.
For all cases, [001] orientated substrates are considered, and the 
out-of-plane lattice constant is chosen to conserve the experimental 
SrFeO$_3$ volume after fixing the in-plane lattice parameters to match 
the following substrates:
SrTiO$_3$ (STO), LSAT$\, = ($LaAlO$_3)_{0.3}($Sr$_2$AlTaO$_6)_{0.7}$,
LaAlO$_3$ (LAO), and YaAlO$_3$ (YAO).
%
}
\end{figure}

In Figure \ref{fig:sfo_strain} we plot the calculated mode 
frequencies of the octahedral rotational patterns for a range of strain 
values corresponding to typical substrates. 
Real frequencies indicate that the cubic lattice is stable and that the 
mode does not spontaneously condense. 
Importantly, we find that all the frequencies are real for 
strain corresponding to coherency on SrTiO$_3$, indicating that the
rotations in our SrFeO$_3$/SrTiO$_3$ heterostructures require the
actual presence of the heterointerface, not only its associated strain.
Under $>1$\% compressive strain, we find an unstable AFD 
rotation which has been reported previously for the case of 
LaAlO$_3$ \cite{Hatt/Spaldin:2008} and provides a low energy route to 
reducing the in-plane lattice parameters without significantly shortening 
the bond lengths.
Interestingly, the $a^0a^0c^+$  tilt is less sensitive to strain and is 
likely due to the rectangular planar coordination of the Sr-site, which 
leads to less energy stabilization from strain-induced covalency 
modifications over the distorted tetragonal coordination available in the 
$a^0a^0c^-$  tilt pattern \cite{Woodward:1997a}.
The $a^-a^0c^0$  AFD tilt also softens with increasing tensile strain to 
accommodate the reduction in the out-of-plane lattice constant as the 
in-plane lattice parameters are elongated.
[The Jahn-Teller and breathing distortions (data not shown) 
remain metallic and are disfavored for all strain states.]
Finally, since we define strain relative to the LSDA equilibrium 
volumes in our calculations, we examine the dependence of the 
SrFeO$_3$ lattice instabilities on unit cell volume. 
We find that all phonon frequencies are real (indicating no instabilities)
for lattice constants within $\pm$2\% of the experimental value 
($a_{0}=3.851$~{\AA}).
Also, the choice of $U$ does not significantly alter the phonon 
dispersions. 
We now discuss how appropriate choice of the antiferrodistortive 
octahedral rotations in the substrate may be used to 
select particular latent lattice instabilities in a functional thin film.
As we showed earlier, if the octahedral tilt pattern in the film 
can be modified, it may be possible to switch between the types 
of electronically-driven structural instabilities that couple to the 
rotational modes.
\begin{figure}
\centering
\includegraphics[width=0.42\textwidth]{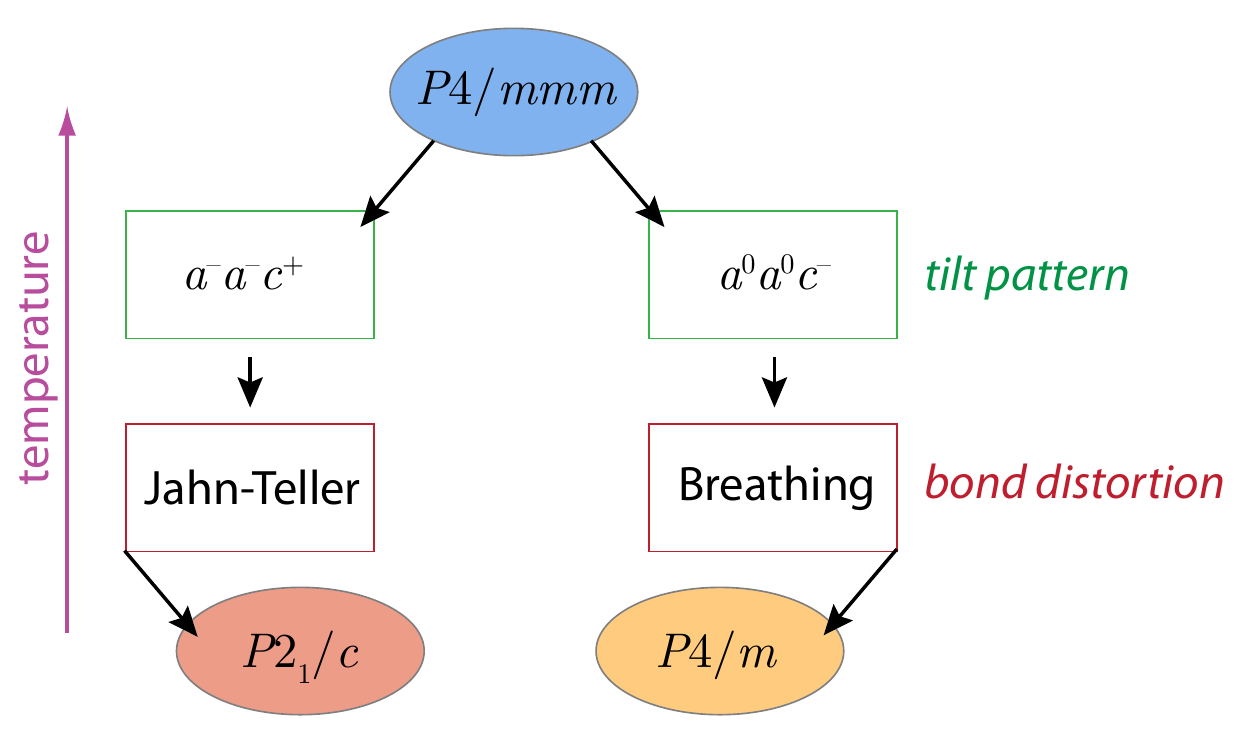}
\caption{\label{fig:sl_transition}
Illustration of how the same high symmetry reference superlattice can support 
either the electronically-driven Jahn-Teller or breathing distortions by 
the condensation of different symmetry octahedral modes with temperature.
The choice of a substrate based on the octahedral rotational properties 
may substitute for temperature to control such phases under standard 
temperatures.
}
\end{figure}
This is shown schematically in Figure \ref{fig:sl_transition} for the 
(SrTiO$_3$)$_n$/(SrFeO$_3$)$_1$ superlattices.
When the superlattice exhibits the orthorhombic $Pnma$ octahedral tilt pattern 
$a^-a^-c^+$, the Jahn-Teller distortion is activated due to the orbital 
degeneracy in the highly confined SrFeO$_3$ layers.
On the other hand, the $a^0a^0c^-$ tilt pattern is found to occur with the 
octahedral breathing distortion.
The experimental situation likely depends on the different temperature scales 
associated with the octahedral and electronic-driven structural distortions; 
since the rotations condense at higher temperatures, particular substrates 
may be chosen to be compatible with certain symmetry 
electronic-instabilities that condense at lower temperatures.
In the case of these ferrate/titanate superlattices,  if a substrate 
that shows orthorhombic or tetragonal octahedral rotations is used, it may be 
possible to strictly select  orbitally- or charge-ordered 
electronic configurations in the ferrate layers by heteroepitaxy.
An interesting question is whether such strong coupling between the 
octahedral tilt modes and electronic instabilities is 
present in bulk perovskite oxides.
Such coupling is active in the isoelectronic $d^4$ mangantite 
compound LaMnO$_3$, where the orthorhombic 
$a^-a^-c^+$ tilt pattern combines with a long-range ordering of Jahn-Teller 
distortions, in preference to the breathing distortion, to induce a metal-insulator phase transition \cite{Carpenter/Howard:2009b}.
The octahedral rotations and La displacements are critical to stabilizing the ground state electronic configuration and the metal-insulator 
transition \cite{Mizokawa/Khomskii/Sawatsky:1999,Pavarini/Koch:2010}.
In fact, without the tilting the $t_{2g}$ and $e_g$ states are forbidden 
to mix by symmetry and the orbitally ordered ground state 
imposed by the Jahn-Teller distortion does not occur \cite{Solovyev/Hamada/Terakura:1996}.
Since orbitally-degenerate perovskite oxides support strong coupling 
between lattice modes of different origins, growth 
on substrates with different octahedral instabilities offers a new 
route to tune the electronic properties of these materials.
In summary we used density functional calculations and the model 
SrFeO$_3$/SrTiO$_3$ superlattice system 
to show that thin film perovskite heterointerfaces  
are not only sensitive to the elastic strain at the interface, but also to 
the octahedral rotation patterns present in the substrate.
We found that the $a^0a^0c^-$ tilt pattern in the SrTiO$_3$ substrate 
propagates into the SrFeO$_3$ film interface layers.
Interestingly, we find that the substrate not only affects the rotation 
patterns of the film, but that the entire heterostructure can adopt
a tilt pattern that is different from that of the parent substrate or
film structure; 
this may partly explain the recent anomalous 
structural phase transitions observed in SrTiO$_3$ substrates used to 
support manganite/cuprate superlattices \cite{Hoppler/Willmott:2008}.
%
%
We point out that the SrTiO$_3$ $a^0a^0c^-$ tilt pattern studied in this 
work is somewhat weakly correlated along the $c$-direction 
and is the reason for the short penetration depth across the interface.
Stronger coupling of substrate coherency-induced octahedral rotations can 
be obtained by choosing substrates which show in-plane rather 
than out-of-plane octahedral rotations patterns with orthorhombic or monoclinic 
space groups.
Alternatively, control of the substrate vicinal cut may be used to control 
the types of octahedral rotational antiphase domain boundaries induced 
by the substrate.
The ability to control octahedral rotation textures present in thin film 
via substrate proximity effects affords an additional parameter for tuning
the atomic and electronic structure of functional oxide films.

\begin{acknowledgments}
This work was supported by a DOD NDSEG Fellowship (JMR) and the National Science
Foundation through grant no.\ DMR 0940420 (NAS). We thank S.\ May, 
C.\ Adamo, and D.\ Schlom for useful discussions.
%
\end{acknowledgments}


\begin{thebibliography}{29}
\expandafter\ifx\csname natexlab\endcsname\relax\def\natexlab#1{#1}\fi
\expandafter\ifx\csname bibnamefont\endcsname\relax
  \def\bibnamefont#1{#1}\fi
\expandafter\ifx\csname bibfnamefont\endcsname\relax
  \def\bibfnamefont#1{#1}\fi
\expandafter\ifx\csname citenamefont\endcsname\relax
  \def\citenamefont#1{#1}\fi
\expandafter\ifx\csname url\endcsname\relax
  \def\url#1{\texttt{#1}}\fi
\expandafter\ifx\csname urlprefix\endcsname\relax\def\urlprefix{URL }\fi
\providecommand{\bibinfo}[2]{#2}
\providecommand{\eprint}[2][]{\url{#2}}

\bibitem[{\citenamefont{Schlom et~al.}(2007)\citenamefont{Schlom, Chen, Eom,
  Rabe, Streiffer, and Triscone}}]{Schlom/Triscone_et_al:2007}
\bibinfo{author}{\bibfnamefont{D.~G.} \bibnamefont{Schlom}},
  \bibinfo{author}{\bibfnamefont{L.-Q.} \bibnamefont{Chen}},
  \bibinfo{author}{\bibfnamefont{C.-B.} \bibnamefont{Eom}},
  \bibinfo{author}{\bibfnamefont{K.~M.} \bibnamefont{Rabe}},
  \bibinfo{author}{\bibfnamefont{S.~K.} \bibnamefont{Streiffer}},
  \bibnamefont{and} \bibinfo{author}{\bibfnamefont{J.-M.}
  \bibnamefont{Triscone}}, \bibinfo{journal}{Ann. Rev. Mater. Res.}
  \textbf{\bibinfo{volume}{37}}, \bibinfo{pages}{589} (\bibinfo{year}{2007}).

\bibitem[{\citenamefont{He et~al.}(2004)\citenamefont{He, Wells, Ban, Alpay,
  Grenier, Shapiro, Si, Clark, and Xi}}]{He/Xi_et_al:2004}
\bibinfo{author}{\bibfnamefont{F.}~\bibnamefont{He}},
  \bibinfo{author}{\bibfnamefont{B.~O.} \bibnamefont{Wells}},
  \bibinfo{author}{\bibfnamefont{Z.-G.} \bibnamefont{Ban}},
  \bibinfo{author}{\bibfnamefont{S.~P.} \bibnamefont{Alpay}},
  \bibinfo{author}{\bibfnamefont{S.}~\bibnamefont{Grenier}},
  \bibinfo{author}{\bibfnamefont{S.~M.} \bibnamefont{Shapiro}},
  \bibinfo{author}{\bibfnamefont{W.}~\bibnamefont{Si}},
  \bibinfo{author}{\bibfnamefont{A.}~\bibnamefont{Clark}}, \bibnamefont{and}
  \bibinfo{author}{\bibfnamefont{X.~X.} \bibnamefont{Xi}},
  \bibinfo{journal}{Phys. Rev. B} \textbf{\bibinfo{volume}{70}},
  \bibinfo{pages}{235405} (\bibinfo{year}{2004}).

\bibitem[{\citenamefont{Xie et~al.}(2008)\citenamefont{Xie, Budnick, Hines,
  Wells, and Woicik}}]{Xie/Woicik_et_al:2008}
\bibinfo{author}{\bibfnamefont{C.~K.} \bibnamefont{Xie}},
  \bibinfo{author}{\bibfnamefont{J.~I.} \bibnamefont{Budnick}},
  \bibinfo{author}{\bibfnamefont{W.~A.} \bibnamefont{Hines}},
  \bibinfo{author}{\bibfnamefont{B.~O.} \bibnamefont{Wells}}, \bibnamefont{and}
  \bibinfo{author}{\bibfnamefont{J.~C.} \bibnamefont{Woicik}},
  \bibinfo{journal}{Appl. Phys. Lett.} \textbf{\bibinfo{volume}{93}},
  \bibinfo{eid}{182507} (\bibinfo{year}{2008}).

\bibitem[{\citenamefont{Hoppler et~al.}(2008)\citenamefont{Hoppler, Stahn,
  Bouyanfif, Malik, Patterson, Willmott, Cristiani, Habermeier, and
  Bernhard}}]{Hoppler/Willmott:2008}
\bibinfo{author}{\bibfnamefont{J.}~\bibnamefont{Hoppler}},
  \bibinfo{author}{\bibfnamefont{J.}~\bibnamefont{Stahn}},
  \bibinfo{author}{\bibfnamefont{H.}~\bibnamefont{Bouyanfif}},
  \bibinfo{author}{\bibfnamefont{V.~K.} \bibnamefont{Malik}},
  \bibinfo{author}{\bibfnamefont{B.~D.} \bibnamefont{Patterson}},
  \bibinfo{author}{\bibfnamefont{P.~R.} \bibnamefont{Willmott}},
  \bibinfo{author}{\bibfnamefont{G.}~\bibnamefont{Cristiani}},
  \bibinfo{author}{\bibfnamefont{H.-U.} \bibnamefont{Habermeier}},
  \bibnamefont{and} \bibinfo{author}{\bibfnamefont{C.}~\bibnamefont{Bernhard}},
  \bibinfo{journal}{Phys. Rev. B} \textbf{\bibinfo{volume}{78}},
  \bibinfo{pages}{134111} (\bibinfo{year}{2008}).

\bibitem[{\citenamefont{Loetzsch et~al.}(2010)\citenamefont{Loetzsch,
  L\"{u}bcke, Uschmann, F\"{o}rster, e, Thuerk, Koettig, Schmidl, and
  Seidel}}]{Loetzsch/Seidel:2010}
\bibinfo{author}{\bibfnamefont{R.}~\bibnamefont{Loetzsch}},
  \bibinfo{author}{\bibfnamefont{A.}~\bibnamefont{L\"{u}bcke}},
  \bibinfo{author}{\bibfnamefont{I.}~\bibnamefont{Uschmann}},
  \bibinfo{author}{\bibfnamefont{E.}~\bibnamefont{F\"{o}rster}},
  \bibinfo{author}{\bibfnamefont{V.~G.} \bibnamefont{e}},
  \bibinfo{author}{\bibfnamefont{M.}~\bibnamefont{Thuerk}},
  \bibinfo{author}{\bibfnamefont{T.}~\bibnamefont{Koettig}},
  \bibinfo{author}{\bibfnamefont{F.}~\bibnamefont{Schmidl}}, \bibnamefont{and}
  \bibinfo{author}{\bibfnamefont{P.}~\bibnamefont{Seidel}},
  \bibinfo{journal}{Appl. Phys. Lett.}  (\bibinfo{year}{2010}).

\bibitem[{\citenamefont{Okamoto et~al.}(2006)\citenamefont{Okamoto, Millis, and
  Spaldin}}]{Okamoto/Millis/Spaldin:2006}
\bibinfo{author}{\bibfnamefont{S.}~\bibnamefont{Okamoto}},
  \bibinfo{author}{\bibfnamefont{A.~J.} \bibnamefont{Millis}},
  \bibnamefont{and} \bibinfo{author}{\bibfnamefont{N.~A.}
  \bibnamefont{Spaldin}}, \bibinfo{journal}{Phys. Rev. Lett.}
  \textbf{\bibinfo{volume}{97}}, \bibinfo{pages}{056802}
  (\bibinfo{year}{2006}).

\bibitem[{\citenamefont{Bhattacharjee et~al.}(2009)\citenamefont{Bhattacharjee,
  Bousquet, and Ghosez}}]{Bhattacharjee/Bousquet/Ghosez:2009}
\bibinfo{author}{\bibfnamefont{S.}~\bibnamefont{Bhattacharjee}},
  \bibinfo{author}{\bibfnamefont{E.}~\bibnamefont{Bousquet}}, \bibnamefont{and}
  \bibinfo{author}{\bibfnamefont{P.}~\bibnamefont{Ghosez}},
  \bibinfo{journal}{Phys. Rev. Lett.} \textbf{\bibinfo{volume}{102}},
  \bibinfo{pages}{117602} (\bibinfo{year}{2009}).

\bibitem[{\citenamefont{Rondinelli et~al.}(2009)\citenamefont{Rondinelli,
  Eidelson, and Spaldin}}]{Rondinelli/Eidelson/Spaldin:2009}
\bibinfo{author}{\bibfnamefont{J.~M.} \bibnamefont{Rondinelli}},
  \bibinfo{author}{\bibfnamefont{A.~S.} \bibnamefont{Eidelson}},
  \bibnamefont{and} \bibinfo{author}{\bibfnamefont{N.~A.}
  \bibnamefont{Spaldin}}, \bibinfo{journal}{Phys. Rev. B}
  \textbf{\bibinfo{volume}{79}}, \bibinfo{pages}{205119}
  (\bibinfo{year}{2009}).

\bibitem[{\citenamefont{Bousquet et~al.}(2008)\citenamefont{Bousquet, Dawber,
  Stucki, Lichtensteiger, Hermet, Gariglio, Triscone, and
  Ghosez}}]{Bousquet/Ghosez_et_al:2008}
\bibinfo{author}{\bibfnamefont{E.}~\bibnamefont{Bousquet}},
  \bibinfo{author}{\bibfnamefont{M.}~\bibnamefont{Dawber}},
  \bibinfo{author}{\bibfnamefont{N.}~\bibnamefont{Stucki}},
  \bibinfo{author}{\bibfnamefont{C.}~\bibnamefont{Lichtensteiger}},
  \bibinfo{author}{\bibfnamefont{P.}~\bibnamefont{Hermet}},
  \bibinfo{author}{\bibfnamefont{S.}~\bibnamefont{Gariglio}},
  \bibinfo{author}{\bibfnamefont{J.-M.} \bibnamefont{Triscone}},
  \bibnamefont{and} \bibinfo{author}{\bibfnamefont{P.}~\bibnamefont{Ghosez}},
  \bibinfo{journal}{Nature} \textbf{\bibinfo{volume}{452}},
  \bibinfo{pages}{732} (\bibinfo{year}{2008}).

\bibitem[{\citenamefont{Zeches et~al.}(2009)\citenamefont{Zeches, Rossell,
  Zhang, Hatt, He, Yang, Kumar, Wang, Melville, Adamo
  et~al.}}]{Hatt/Ramesh:2009}
\bibinfo{author}{\bibfnamefont{R.~J.} \bibnamefont{Zeches}},
  \bibinfo{author}{\bibfnamefont{M.~D.} \bibnamefont{Rossell}},
  \bibinfo{author}{\bibfnamefont{J.~X.} \bibnamefont{Zhang}},
  \bibinfo{author}{\bibfnamefont{A.~J.} \bibnamefont{Hatt}},
  \bibinfo{author}{\bibfnamefont{Q.}~\bibnamefont{He}},
  \bibinfo{author}{\bibfnamefont{C.-H.} \bibnamefont{Yang}},
  \bibinfo{author}{\bibfnamefont{A.}~\bibnamefont{Kumar}},
  \bibinfo{author}{\bibfnamefont{C.~H.} \bibnamefont{Wang}},
  \bibinfo{author}{\bibfnamefont{A.}~\bibnamefont{Melville}},
  \bibinfo{author}{\bibfnamefont{C.}~\bibnamefont{Adamo}},
  \bibnamefont{et~al.}, \bibinfo{journal}{Science}
  \textbf{\bibinfo{volume}{326}}, \bibinfo{pages}{977} (\bibinfo{year}{2009}).

\bibitem[{\citenamefont{Jia et~al.}(2009)\citenamefont{Jia, Mi, Faley, Poppe,
  Schubert, and Urban}}]{Jia/Urban_et_al:2009}
\bibinfo{author}{\bibfnamefont{C.~L.} \bibnamefont{Jia}},
  \bibinfo{author}{\bibfnamefont{S.~B.} \bibnamefont{Mi}},
  \bibinfo{author}{\bibfnamefont{M.}~\bibnamefont{Faley}},
  \bibinfo{author}{\bibfnamefont{U.}~\bibnamefont{Poppe}},
  \bibinfo{author}{\bibfnamefont{J.}~\bibnamefont{Schubert}}, \bibnamefont{and}
  \bibinfo{author}{\bibfnamefont{K.}~\bibnamefont{Urban}},
  \bibinfo{journal}{Phys. Rev. B} \textbf{\bibinfo{volume}{79}},
  \bibinfo{pages}{081405} (\bibinfo{year}{2009}).

\bibitem[{\citenamefont{{May} et~al.}(2010)\citenamefont{{May}, {Kim},
  {Rondinelli}, {Karapetrova}, {Spaldin}, {Bhattacharya}, and
  {Ryan}}}]{May/Rondinelli:2010}
\bibinfo{author}{\bibfnamefont{S.~J.} \bibnamefont{{May}}},
  \bibinfo{author}{\bibfnamefont{J.}~\bibnamefont{{Kim}}},
  \bibinfo{author}{\bibfnamefont{J.~M.} \bibnamefont{{Rondinelli}}},
  \bibinfo{author}{\bibfnamefont{E.}~\bibnamefont{{Karapetrova}}},
  \bibinfo{author}{\bibfnamefont{N.~A.} \bibnamefont{{Spaldin}}},
  \bibinfo{author}{\bibfnamefont{A.}~\bibnamefont{{Bhattacharya}}},
  \bibnamefont{and} \bibinfo{author}{\bibfnamefont{P.~J.}
  \bibnamefont{{Ryan}}}, \bibinfo{journal}{ArXiv e-prints}
  (\bibinfo{year}{2010}), \eprint{1002.1317}.

\bibitem[{\citenamefont{MacChesney et~al.}(1965)\citenamefont{MacChesney,
  Sherwood, and Potter}}]{Macchesney/Potter_et_al:1965}
\bibinfo{author}{\bibfnamefont{J.~B.} \bibnamefont{MacChesney}},
  \bibinfo{author}{\bibfnamefont{R.~C.} \bibnamefont{Sherwood}},
  \bibnamefont{and} \bibinfo{author}{\bibfnamefont{J.~F.}
  \bibnamefont{Potter}}, \bibinfo{journal}{J.\ Chem. Phys.}
  \textbf{\bibinfo{volume}{43}}, \bibinfo{pages}{1907} (\bibinfo{year}{1965}).

\bibitem[{\citenamefont{Jauch and Palmer}(1999)}]{Jauch/Palmer:1999}
\bibinfo{author}{\bibfnamefont{W.}~\bibnamefont{Jauch}} \bibnamefont{and}
  \bibinfo{author}{\bibfnamefont{A.}~\bibnamefont{Palmer}},
  \bibinfo{journal}{Phys. Rev. B} \textbf{\bibinfo{volume}{60}},
  \bibinfo{pages}{2961} (\bibinfo{year}{1999}).

\bibitem[{\citenamefont{Kresse and
  Furthm{\"u}ller}(1996)}]{Kresse/Furthmueller_PRB:1996}
\bibinfo{author}{\bibfnamefont{G.}~\bibnamefont{Kresse}} \bibnamefont{and}
  \bibinfo{author}{\bibfnamefont{J.}~\bibnamefont{Furthm{\"u}ller}},
  \bibinfo{journal}{Phys. Rev. B} \textbf{\bibinfo{volume}{54}},
  \bibinfo{pages}{11169} (\bibinfo{year}{1996}).

\bibitem[{\citenamefont{Kresse and Joubert}(1999)}]{Kresse/Joubert:1999}
\bibinfo{author}{\bibfnamefont{G.}~\bibnamefont{Kresse}} \bibnamefont{and}
  \bibinfo{author}{\bibfnamefont{D.}~\bibnamefont{Joubert}},
  \bibinfo{journal}{Phys. Rev. B} \textbf{\bibinfo{volume}{59}},
  \bibinfo{pages}{1758} (\bibinfo{year}{1999}).

\bibitem[{\citenamefont{Dudarev et~al.}(1998)\citenamefont{Dudarev, Botton,
  Savrasov, Humphreys, and Sutton}}]{Dudarev_et_al:1998}
\bibinfo{author}{\bibfnamefont{S.~L.} \bibnamefont{Dudarev}},
  \bibinfo{author}{\bibfnamefont{G.~A.} \bibnamefont{Botton}},
  \bibinfo{author}{\bibfnamefont{S.~Y.} \bibnamefont{Savrasov}},
  \bibinfo{author}{\bibfnamefont{C.~J.} \bibnamefont{Humphreys}},
  \bibnamefont{and} \bibinfo{author}{\bibfnamefont{A.~P.}
  \bibnamefont{Sutton}}, \bibinfo{journal}{Phys. Rev. B}
  \textbf{\bibinfo{volume}{57}}, \bibinfo{pages}{1505} (\bibinfo{year}{1998}).

\bibitem[{\citenamefont{Shein et~al.}(2005)\citenamefont{Shein, Shein,
  Kozhevnikov, and Ivanovskii}}]{Shein/Ivanovskii_et_al:2005}
\bibinfo{author}{\bibfnamefont{I.}~\bibnamefont{Shein}},
  \bibinfo{author}{\bibfnamefont{K.}~\bibnamefont{Shein}},
  \bibinfo{author}{\bibfnamefont{V.}~\bibnamefont{Kozhevnikov}},
  \bibnamefont{and}
  \bibinfo{author}{\bibfnamefont{A.}~\bibnamefont{Ivanovskii}},
  \bibinfo{journal}{Phys. Sol. State} \textbf{\bibinfo{volume}{47}},
  \bibinfo{pages}{2082} (\bibinfo{year}{2005}).

\bibitem[{\citenamefont{Rondinelli and
  Spaldin}(2010)}]{Rondinelli/Spaldin:2010}
\bibinfo{author}{\bibfnamefont{J.~M.} \bibnamefont{Rondinelli}}
  \bibnamefont{and} \bibinfo{author}{\bibfnamefont{N.~A.}
  \bibnamefont{Spaldin}}, \bibinfo{journal}{Phys. Rev. B}
  \textbf{\bibinfo{volume}{81}}, \bibinfo{pages}{085109}
  (\bibinfo{year}{2010}).

\bibitem[{\citenamefont{Unoki and Sakudo}(1967)}]{Unoki/Sakudo:1967}
\bibinfo{author}{\bibfnamefont{H.}~\bibnamefont{Unoki}} \bibnamefont{and}
  \bibinfo{author}{\bibfnamefont{T.}~\bibnamefont{Sakudo}},
  \bibinfo{journal}{J. Phys. Soc. Jap.} \textbf{\bibinfo{volume}{23}},
  \bibinfo{pages}{546} (\bibinfo{year}{1967}).

\bibitem[{\citenamefont{Sai and Vanderbilt}(2000)}]{Sai/Vanderbilt:2000}
\bibinfo{author}{\bibfnamefont{N.}~\bibnamefont{Sai}} \bibnamefont{and}
  \bibinfo{author}{\bibfnamefont{D.}~\bibnamefont{Vanderbilt}},
  \bibinfo{journal}{Phys. Rev. B} \textbf{\bibinfo{volume}{62}},
  \bibinfo{pages}{13942} (\bibinfo{year}{2000}).

\bibitem[{\citenamefont{Lin et~al.}(2006)\citenamefont{Lin, Huang, and
  Guo}}]{Lin/Guo:2006}
\bibinfo{author}{\bibfnamefont{C.-H.} \bibnamefont{Lin}},
  \bibinfo{author}{\bibfnamefont{C.-M.} \bibnamefont{Huang}}, \bibnamefont{and}
  \bibinfo{author}{\bibfnamefont{G.~Y.} \bibnamefont{Guo}},
  \bibinfo{journal}{J. Appl. Phys.} \textbf{\bibinfo{volume}{100}},
  \bibinfo{eid}{084104} (\bibinfo{year}{2006}).

\bibitem[{\citenamefont{{Hatt} and {Spaldin}}(2008)}]{Hatt/Spaldin:2008}
\bibinfo{author}{\bibfnamefont{A.~J.} \bibnamefont{{Hatt}}} \bibnamefont{and}
  \bibinfo{author}{\bibfnamefont{N.~A.} \bibnamefont{{Spaldin}}},
  \bibinfo{journal}{ArXiv e-prints}  (\bibinfo{year}{2008}),
  \eprint{0808.3792}.

\bibitem[{\citenamefont{Woodward}(1997)}]{Woodward:1997a}
\bibinfo{author}{\bibfnamefont{P.}~\bibnamefont{Woodward}},
  \bibinfo{journal}{Acta Cryst} \textbf{\bibinfo{volume}{B53}},
  \bibinfo{pages}{44} (\bibinfo{year}{1997}).

\bibitem[{\citenamefont{Carpenter and Howard}(2009)}]{Carpenter/Howard:2009b}
\bibinfo{author}{\bibfnamefont{M.~A.} \bibnamefont{Carpenter}}
  \bibnamefont{and} \bibinfo{author}{\bibfnamefont{C.~J.}
  \bibnamefont{Howard}}, \bibinfo{journal}{Acta Cryst. B}
  \textbf{\bibinfo{volume}{65}}, \bibinfo{pages}{147} (\bibinfo{year}{2009}).

\bibitem[{\citenamefont{Mizokawa et~al.}(1999)\citenamefont{Mizokawa, Khomskii,
  and Sawatzky}}]{Mizokawa/Khomskii/Sawatsky:1999}
\bibinfo{author}{\bibfnamefont{T.}~\bibnamefont{Mizokawa}},
  \bibinfo{author}{\bibfnamefont{D.~I.} \bibnamefont{Khomskii}},
  \bibnamefont{and} \bibinfo{author}{\bibfnamefont{G.~A.}
  \bibnamefont{Sawatzky}}, \bibinfo{journal}{Phys. Rev. B}
  \textbf{\bibinfo{volume}{60}}, \bibinfo{pages}{7309} (\bibinfo{year}{1999}).

\bibitem[{\citenamefont{Pavarini and Koch}(2010)}]{Pavarini/Koch:2010}
\bibinfo{author}{\bibfnamefont{E.}~\bibnamefont{Pavarini}} \bibnamefont{and}
  \bibinfo{author}{\bibfnamefont{E.}~\bibnamefont{Koch}},
  \bibinfo{journal}{Phys. Rev. Lett.} \textbf{\bibinfo{volume}{104}},
  \bibinfo{pages}{086402} (\bibinfo{year}{2010}).

\bibitem[{\citenamefont{Solovyev et~al.}(1996)\citenamefont{Solovyev, Hamada,
  and Terakura}}]{Solovyev/Hamada/Terakura:1996}
\bibinfo{author}{\bibfnamefont{I.}~\bibnamefont{Solovyev}},
  \bibinfo{author}{\bibfnamefont{N.}~\bibnamefont{Hamada}}, \bibnamefont{and}
  \bibinfo{author}{\bibfnamefont{K.}~\bibnamefont{Terakura}},
  \bibinfo{journal}{Phys. Rev. Lett.} \textbf{\bibinfo{volume}{76}},
  \bibinfo{pages}{4825} (\bibinfo{year}{1996}).

\bibitem[{\citenamefont{Courtens}(1972)}]{Courtens:1972}
\bibinfo{author}{\bibfnamefont{E.}~\bibnamefont{Courtens}},
  \bibinfo{journal}{Phys. Rev. Lett.} \textbf{\bibinfo{volume}{29}},
  \bibinfo{pages}{1380} (\bibinfo{year}{1972}).

\end{thebibliography}

\end{document}